\documentclass[letterpaper,aps,prb,preprint,superscriptaddress]{revtex4}
\usepackage{amssymb}
\usepackage{color}
\usepackage[colorlinks=true,citecolor=blue,linkcolor=blue]{hyperref}
\usepackage{natbib}
\usepackage{setspace}
\usepackage{graphicx}
\usepackage{mhchem}
\usepackage{tabularx}
\usepackage[english]{babel}

\definecolor{Red}{RGB}{200,26,28}
\usepackage{siunitx}
\DeclareSIUnit{\molar}{M}
\DeclareSIUnit{\GO}{\ensuremath{\text{G}_0}}
\begin{document}
\title{Controlled quantum dot formation in atomically engineered graphene nanoribbons field-effect transistors}

\author{Maria El Abbassi}
\email{m.elabbassi-1@tudelft.nl}
\affiliation{Department of Physics, University of Basel, Klingelbergstrasse 82, CH-4056 Basel, Switzerland}
\affiliation{Empa, Swiss Federal Laboratories for Materials Science and Technology, Transport at nanoscale interfaces Laboratory, CH-8600 D\"{u}bendorf, Switzerland.}
\affiliation{Kavli Institute of Nanoscience, Delft University of Technology, Lorentzweg 1, 2628 CJ Delft, The Netherlands}
\affiliation{These authors contributed equally}

\author{Mickael Perrin}
\affiliation{Empa, Swiss Federal Laboratories for Materials Science and Technology, Transport at nanoscale interfaces Laboratory, CH-8600 D\"{u}bendorf, Switzerland.}
\affiliation{These authors contributed equally}

\author{Gabriela Borin Barin}
\affiliation{Empa, Swiss Federal Laboratories for Materials Science and Technology, nanotech@surfaces Laboratory, CH-8600 D\"{u}bendorf, Switzerland.}

\author{Sara Sangtarash}
\affiliation{Department of Physics, Lancaster University, Lancaster LA1 4YB, United Kingdom}
\affiliation{School of Engineering, University of Warwick, Coventry, CV4 7AL, United Kingdom}

\author{Jan Overbeck}
\affiliation{Department of Physics, University of Basel, Klingelbergstrasse 82, CH-4056 Basel, Switzerland}
\affiliation{Empa, Swiss Federal Laboratories for Materials Science and Technology, Transport at nanoscale interfaces Laboratory, CH-8600 D\"{u}bendorf, Switzerland.}

\author{Oliver Braun}
\affiliation{Department of Physics, University of Basel, Klingelbergstrasse 82, CH-4056 Basel, Switzerland}
\affiliation{Empa, Swiss Federal Laboratories for Materials Science and Technology, Transport at nanoscale interfaces Laboratory, CH-8600 D\"{u}bendorf, Switzerland.}

\author{Colin Lambert} 
\affiliation{Department of Physics, Lancaster University, Lancaster LA1 4YB, United Kingdom}

\author{Qiang Sun}
\affiliation{Empa, Swiss Federal Laboratories for Materials Science and Technology, nanotech@surfaces Laboratory, CH-8600 D\"{u}bendorf, Switzerland.}

\author{Thorsten Prechtl}
\affiliation{Max Planck Institute for Polymer Research, Ackermannweg 10, 55128 Mainz, Germany}

\author{Akimitsu Narita}
\affiliation{Max Planck Institute for Polymer Research, Ackermannweg 10, 55128 Mainz, Germany}

\author{Klaus M\"ullen}
\affiliation{Max Planck Institute for Polymer Research, Ackermannweg 10, 55128 Mainz, Germany}

\author{Pascal Ruffieux}
\affiliation{Empa, Swiss Federal Laboratories for Materials Science and Technology, nanotech@surfaces Laboratory, CH-8600 D\"{u}bendorf, Switzerland.}

\author{Hatef Sadeghi}
\email{Hatef.Sadeghi@warwick.ac.uk}
\affiliation{Department of Physics, Lancaster University, Lancaster LA1 4YB, United Kingdom}
\affiliation{School of Engineering, University of Warwick, Coventry, CV4 7AL, United Kingdom}

\author{Roman Fasel}
\email{roman.fasel@empa.ch}
\affiliation{Empa, Swiss Federal Laboratories for Materials Science and Technology, nanotech@surfaces Laboratory, CH-8600 D\"{u}bendorf, Switzerland.}
\affiliation{Department of Chemistry and Biochemistry, University of Bern, Freiestrasse 3, CH-3012 Bern, Switzerland}

\author{Michel Calame}
\email{michel.calame@empa.ch}
\affiliation{Department of Physics, University of Basel, Klingelbergstrasse 82, CH-4056 Basel, Switzerland}
\affiliation{Empa, Swiss Federal Laboratories for Materials Science and Technology, Transport at nanoscale interfaces Laboratory, CH-8600 D\"{u}bendorf, Switzerland.}
\affiliation{Swiss Nanoscience Institute, University of Basel, 4056 Basel, Switzerland.}

\maketitle

\newpage

\textbf{Graphene nanoribbons (GNRs) have attracted a strong interest from researchers worldwide, as they constitute an emerging class of quantum-designed materials. 
The major challenges towards their exploitation in electronic applications include reliable contacting, complicated by their small size ($<$ 50 nm), as well as the preservation of their physical properties upon device integration. 
In this combined experimental and theoretical study, we report on the quantum dot (QD) behavior of atomically precise GNRs integrated in a device geometry. The devices consist of a film of aligned 5-atoms wide GNRs (5-AGNRs) transferred onto graphene electrodes with a sub 5-nm nanogap. We demonstrate that the narrow-bandgap 5-AGNRs exhibit metal-like behavior resulting in linear IV curves for low bias voltages at room temperature and single-electron transistor behavior for temperatures below 150~K. By performing spectroscopy of the molecular levels at 13~K, we obtain addition energies in the range of 200-300 meV. DFT calculations predict comparable addition energies and reveal the presence of two electronic states within the bandgap of infinite ribbons when the finite length of the 5-AGNRs is accounted for. By demonstrating the preservation of the 5-AGNRs electronic properties upon device integration, as demonstrated by transport spectroscopy, our study provides a critical step forward in the realisation of more exotic GNR-based nano-electronic devices. 
 }

\section{Introduction}
Graphene nanoribbons (GNRs) have been shown to host a variety of intriguing physical properties such as half-metallicity, spin-polarized edge states and topologically protected edge states, realized solely by shaping their edge structure with atomic precision.\cite{ch7_GW,ch7_gabi,ch7_Fasel1,ch7_Fasel2, ch7_Cromie}. 
In particular, the chemical tunability of their band structure makes them very appealing objects both for fundamental research as well as functional materials in nanoelectronic devices.  
For example, due to their extremely small size, GNRs have been shown to behave as quantum dots tunable by chemical design\cite{wang2017quantum}, making them highly suited for narrow-band light-emitting devices \cite{Schull}.
In addition, their spin properties have attracted attention as magnetic nanoribbons have been shown to possess long coherence times, offering encouraging prospects for their use in spintronic devices\cite{Lapo}. Finally, the presence of topologically non-trivial quantum phases resulting in zero-energy states \cite{groning2018engineering} offer prospects for electrically tunable single-photon emitters.

Most of these exotic properties, however, have been studied in-situ using scanning probe techniques\cite{Schull,Lapo,li2018survival,Ernst}. The integration of GNRs into electronic devices, and in particular the preservation of their electronic structure, remains challenging and a largely unexplored area. Only a few on-chip measurements have been reported\cite{chen2016synthesis,ch7_Fasel3,candini,preis2019graphene} so far and show that the device properties are critically determined by the quality of the GNRs, as well as the quality of the GNR/electrode interface.  
Recently, the first high-performance field-effect transistors based on wide-bandgap 9- and 13 atoms GNRs have been realized using metallic contacts\cite{ch7_Fasel3}. Charge transport was there shown to be limited by several factors such as the electrode/GNR interface and the quality of the film after transfer. The growth process was also shown to play a role in a study performed by Martini et al.\cite{candini} where UHV-grown 9-AGNRs incorporated between graphene electrodes are claimed to provide higher device performance than CVD-grown 9-AGNRs. This higher quality of UHV-grown GNRs is of particular interest for 5-AGNRs. A recent STM study has indeed demonstrated that very short UHV-grown 5-AGNRs (3-5 units) show a quasi-metallic behavior and possess the smallest HOMO-LUMO gap reported for GNRs measured on surfaces so far ($\approx$100 meV)\cite{ch7_metalic5}. This ultra-small bandgap has not yet been verified in devices, a challenge requiring the preservation of the GNR integrity upon device integration, and a weak electrode/GNR coupling for the reliable extraction of addition energies.

In this work, we demonstrate the formation of quantum dots in GNR-based field-effect transistors. The properties of the quantum dots are determined by the band structure of the GNRs. Our devices, based on 5-AGNR films that were transferred onto sub 5-nm graphene gaps \cite{Prins,ch7_el2017electroburning}, are characterized electrically, both at room and cryogenic temperatures. At room temperature, the devices present a metal-like behavior with a linear low-bias conductance, whereas at low temperature, single-electron tunneling and Coulomb blockade are observed. Remarkably, this quantum dot behavior is preserved up to temperatures as high as 150~K. The values of the HOMO-LUMO gap extracted from transport measurements are comparable with density functional theory calculations and confirm that the electronic properties of the 5-AGNRs are preserved during device integration.

\section{Device integration}
\begin{figure*}[ht!]
\centering \includegraphics[width=\textwidth]{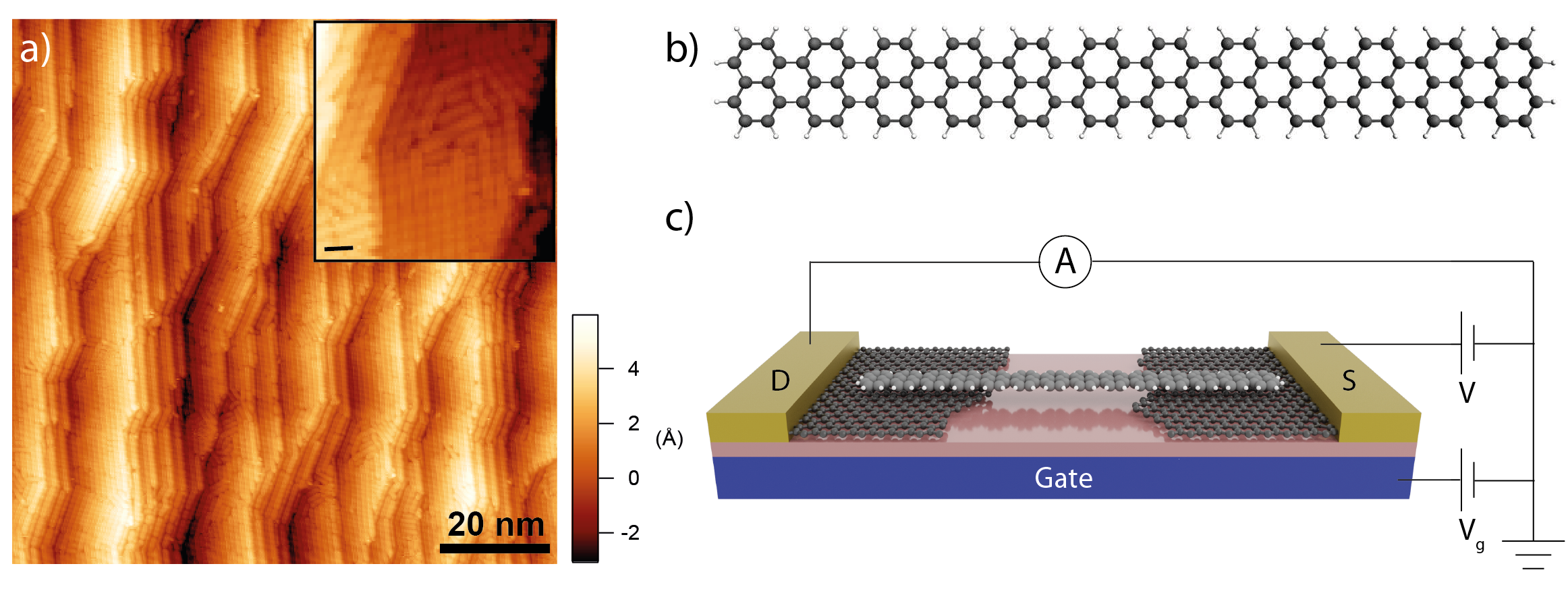}
\caption{\textbf{STM characterization and device geometry} a) STM image of as-synthesized aligned 5-AGNR on a Au (788) surface ($V_s$=-1.5V, $I_t$=0.06nA). b) Schematic of a 5-AGNR. c) Schematic of the 5-AGNR bridging a graphene nanogap and of the electrical transport measurement setup.. }
\label{figure1}
\end{figure*}

The graphene nanogaps used as electrodes to contact the 5-AGNRs are obtained by electrical breakdown of CVD graphene\cite{ch7_nef2014high,ch7_el2017electroburning}. These gaps, being flat and gateable, represent an ideal platform to study the properties of nano-objects \cite{pnas15,anchor0,Prins} and, in particular GNRs. Moreover, the choice of chemical vapor deposition (CVD)-grown graphene allows for fabrication of hundreds of devices in parallel, resulting in large sample statistics. A description of the device fabrication is presented in the Methods section. 

The bottom-up synthesis of aligned 5-AGNRs (schematically represented in Fig.~\ref{figure1}.b) was carried out in ultra-high vacuum via sequential on-surface chemical reactions (details in the method section). The high quality and degree of alignment of 5-AGNRs were verified \textit{in-situ} by scanning tunneling microscopy (STM) at room temperature (see Fig.~\ref{figure1}.a). STM characterization of the 5-AGNRs reveals a distribution of lengths between 2 and 10 nm and shows that they are aligned along the narrow terraces of the Au(788) substrate. 
Controlling the growth direction of the 5-AGNRs allows us to orient them along the source-drain axis of the devices, thereby increasing the yield of GNRs bridging the two electrodes. In order to integrate 5-AGNRs into the graphene junctions, an electrochemical delamination transfer process (or ''bubble transfer'') was employed \cite{ch7_gabi} (details in the method section). A schematic illustration of the device in which a 5-AGNR is bridging a graphene nanogap is given in Fig.~\ref{figure1}c.

\begin{figure*}[ht!]
\centering \includegraphics[width=\textwidth]{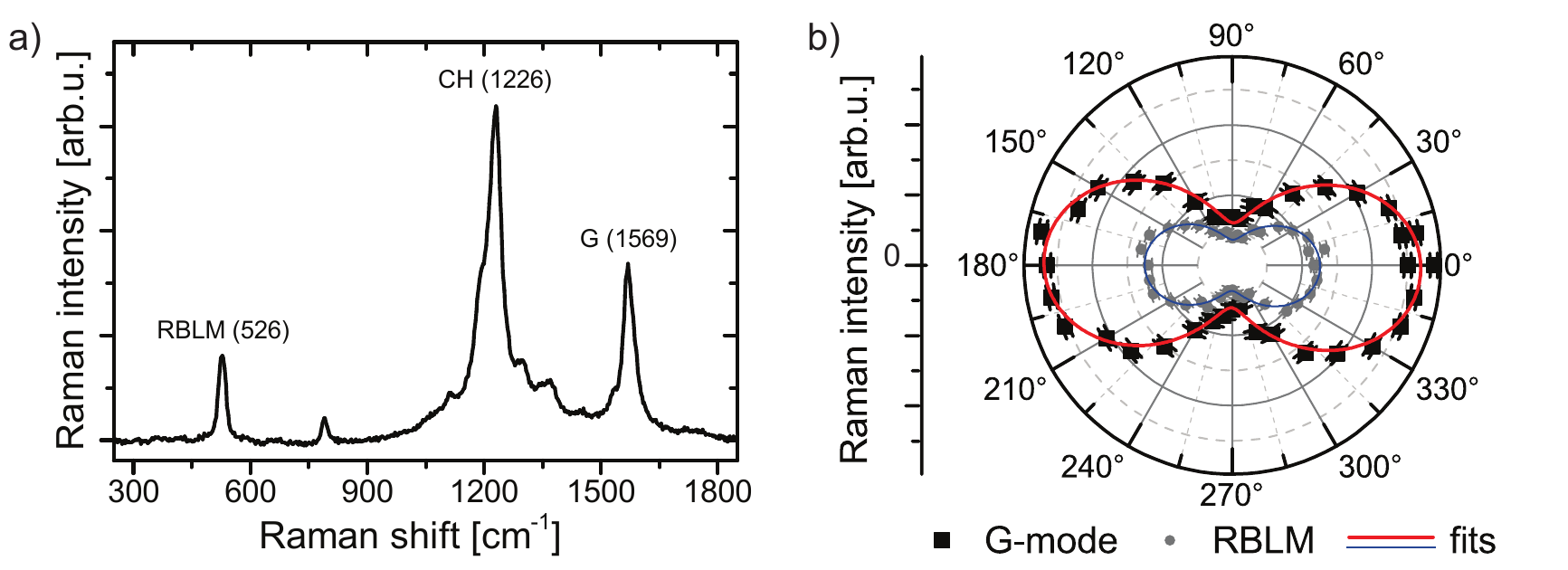}
\caption{\textbf{Raman characterization} a) Raman characterization of the 5-AGNR film acquired after transfer onto the devices. The Raman spectrum was acquired with an excitation wavelength of 785 nm under vacuum conditions. The characteristic radial breathing like mode (RBLM, 526 cm$^{-1} $), an edge-related mode ($\approx$1226 cm$^{-1}$) and the G-mode ($\approx$ 1569 cm$^{-1}$) are clearly visible. b) Polar plot of Raman intensity of the G-mode as a function of polarization axis of the exciting laser relative to the source-drain axis. The intensity shows a clear polarization anisotropy. The data are following the expected cos$^2(\theta)$ dependence (red line). The extracted average ribbon alignment is -1$^\circ$ ($\pm$ 4$^\circ$) degrees with respect to the source-drain axis.}
\label{figure2}
\end{figure*}

\section{Raman characterization}

The graphene nanogap devices with the transferred 5-AGNR film were investigated by Raman spectroscopy to verify the quality and degree of alignment of the GNRs\cite{Overbeck2019,Overbeck2019a}. Figure~\ref{figure2}a presents a Raman spectrum of the transferred 5-AGNR film. The characteristic radial breathing-like-mode (RBLM, 526 cm$^{-1} $), the edge-related mode ($\approx$1226 cm$^{-1}$) and the G-mode ($\approx$ 1569 cm$^{-1}$) are clearly visible \cite{Raman-GNR}. This confirms that the film quality is preserved during the transfer process. Furthermore, these measurements were performed after the electrical characterization and therefore confirm that the film quality is also preserved during measurement procedure. To investigate the alignment of the 5-AGNRs, we have performed polarization-dependent Raman measurements. Fig.~\ref{figure2}b shows the dependence of the Raman intensity of the G-mode on the angle of polarization of the exciting laser. We observe a strong polarization anisotropy with the maximum intensity measured along an axis forming an angle of -1$^\circ$ ($\pm$ 4$^\circ$) with the source-drain axis. This confirms that the ribbons maintain their alignment after transfer and that they are aligned perpendicular to the nanogaps.

\begin{figure*}[ht!]
\centering \includegraphics[width=\textwidth]{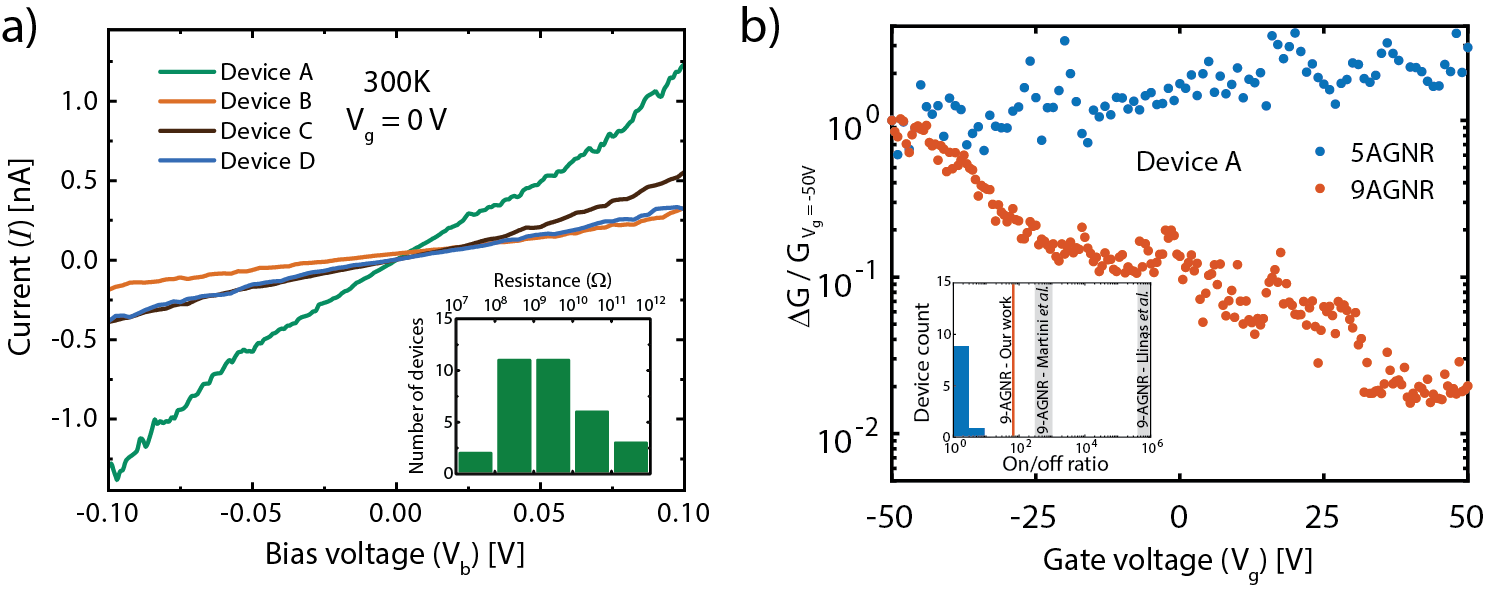}
\caption{ \textbf{Electrical device characterization at room temperature} (a) IV curves measured on 4 devices after transfer of the 5-AGNR at zero gate voltage. The bottom-right inset corresponds to the histograms of the low-bias resistance measured for all the 5-AGNR devices. b) Conductance at fixed bias voltage as a function of gate voltage for the 5-AGNR (device A) and 9-AGNR.} 
\label{figure3}
\end{figure*}

\section{Electrical characterization}


Fig.~\ref{figure3}a presents IV curves recorded at zero gate voltage on 4 typical 5-AGNR devices (label A-D) characterized at room temperature. In the $\pm0.1$V low-bias voltage range, we observe linear IV curves with currents in the nA regime, while the  current is below our detection limit before the GNRs transfer (see Fig.~S1 of the supporting information). This metal-like behavior was observed for all 31 devices investigated at room temperature. 

The bottom-right inset of Fig.~\ref{figure3}a presents a histogram of the low-bias resistance for all devices (fit range $\pm$50~mV). While the linearity of the IV curves is observed for all the devices, a large range of resistances is measured (from 0.01 to 10~G$\Omega$) with an average value around 100 M$\Omega$. 
We attribute this device-to-device variation to possible differences in the number and density of ribbons bridging the gap and the difficulties in controlling the GNR transfer process from the growth substrate to the device substrate. We are currently dedicating substantial efforts to better characterize the GNR properties after transfer\cite{overbeck2019optimized,overbeck2019universal}. We nevertheless can observe that the electronic properties of the 5-AGNRs are preserved once integrated in a device structure. In particular, the formation of quantum dots at low temperature allowed us to extract addition energies reflecting the electronic structure of the 5-AGNRs, as discussed below.

An example of the gate dependence of the conductance value is given in Fig.~\ref{figure3}b for a 5-AGNR device (device A), alongside a measurement on a 9-AGNR device, shown as comparison. While a strong gate dependence is observed for the 9-AGNR with an on/off ratio of about 70, no appreciable dependence is present for the 5-AGNR device. We note that device A has a pronounced gate dependence at 13K (see Fig. 4e). The absence of gate dependence is therefore attributed to the low bandgap of the 5-AGNR, in combination with the thermal smearing of the transmission function at room temperature. An overview of the devices is presented in the inset and contrasted with the measured gate dependence of the 9-AGNR, as well as values on 9-AGNR-based devices reported in literature (with values reaching as high as 10$^5$)\cite{ch7_Fasel3,candini} highlighting the difference to our metal-like GNRs. \\

\begin{figure*}[ht!]
\centering \includegraphics[width=\textwidth]{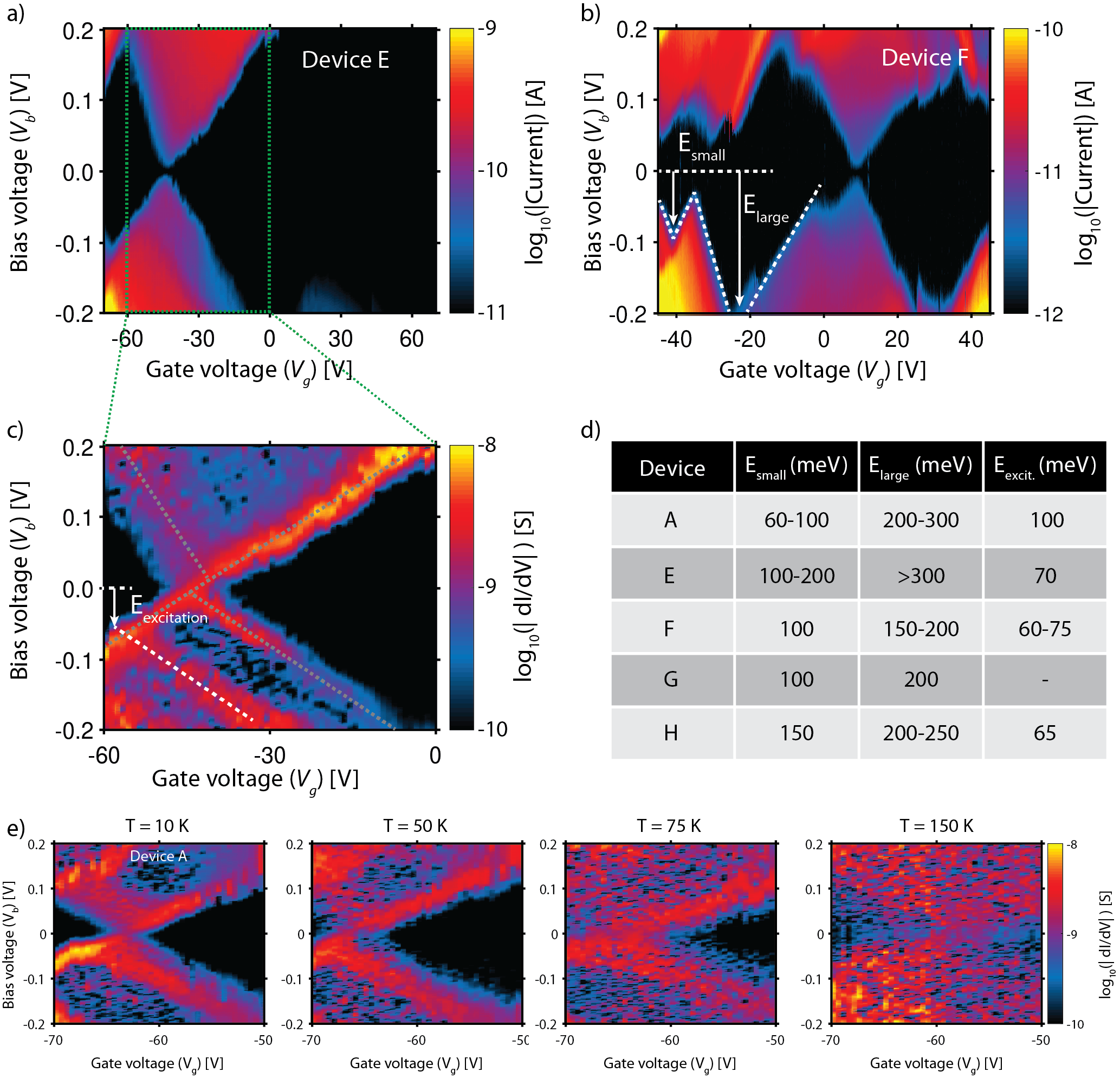}
\caption{ \textbf{Electrical characterization of two devices at 13K} (a,b) Color-coded current maps as a function of gate- and bias voltage for device E and F, respectively (c) Differential conductance (dI/dV, stability diagrams) recorded on device E. The black area corresponds to a blockade regime, while the areas with high currents are in the single-electron tunneling regime. (d) Overview table containing devices A,E-H with the corresponding addition- and excitation energies. e) Stability diagrams recorded on device A at 10K, 50K, 75K and 150K.} 
\label{figure4}
\end{figure*}



To perform a spectroscopic characterization of the energy levels in 5-AGNRs junctions, the devices were cooled down and electrically characterized at 13~K. Figure~\ref{figure4}a and b show color-coded current versus bias- and gate voltage maps (stability diagrams), recorded on devices E and F. The general aspect of the stability diagrams reveal the formation of quantum dots in the GNR devices. In the black regions, no current flows through the device as neither the temperature, nor the bias voltage provide the energy required for adding an extra electron onto the nanoribbon. The edges of this diamond-shaped blocking region (Coulomb blockade regime) correspond to the onset of resonant transport. For bias voltages exceeding the resonance condition, single-electron tunneling occurs (SET regime). 
We extract the energy required to add an electron to the quantum dot as illustrated in Fig.~\ref{figure4}b for two diamonds ($E_{small}$ and $E_{large}$ correspond to half of the height of the Coulomb diamonds). Typically, weakly coupled quantum dots give rise to stability diagrams exhibiting alternating small and larger Coulomb diamonds due to the interplay between the quantum mechanical level spacing and charging energy.\cite{ch7_Herre} We observe such size variations experimentally and, following this logic, we tentatively divided the diamonds in these two categories. We observe characteristic energies in the range 50-300 meV (see table in Fig.~\ref{figure4}d). 

Figure~\ref{figure4}c presents the differential conductance map of device E, zoomed in to the SET region. Here, several resonances are visible (gray dashed lines) which form the edges of the Coulomb diamonds. The edges width depend on the thermal energy (k$_b$T) and the coupling $\Gamma$ of the ribbon to the graphene leads. 
Some of the resonances are asymmetric upon bias polarity inversion, which may be attributed to an asymmetric coupling to the source and drain electrodes as a result of different overlap between the ribbon and the two leads. Also, an additional resonance (white dashed line) running parallel to the diamond edge in the SET regime is observed. This resonance correspond to an electronic or vibrational excited state of the ribbons that creates an additional transport channel. The energy of this excitation can be extracted from the graph by measuring the intersection between the excitation lines and the edge of the Coulomb diamond. The extracted excitation energies are listed in Fig.~\ref{figure4}d for all devices, all lying in the range from 60 to 100 meV. The presence of such excitations is commonly interpreted as a signature of a single quantum dot dominating transport in a particular gate voltage range\cite{ch7_Herre,Mika}. Note that in device F (Fig.~\ref{figure4}b), we can observe partially overlapping diamonds together with the absence of a crossing point at $V_g$=-35 V, suggesting that two or more GNRs contribute to the transport\cite{ch7_Herre,Mika}. In semiconducting quantum dots formed in systems such as 2DEGs, top-down patterned graphene or carbon nanotubes, the addition energies are typically $<$10meV, and several Coulomb diamonds can be observed\cite{jarillo2004electron,thomas2019highly,gustavsson2006counting}. In our case, the addition energies are substantially larger and reach up to several hundreds of millielectronvolts, making the diamond structure apparent up to temperatures as high as 150~K. Fig.~\ref{figure4}e shows four stability diagrams between 10~K and 150~K on device A in the gate voltage range -70~V to -50~V. The plots show a crossing point around -63~V, where the diamond edges meet. Upon increasing the temperature, the resonances running along the diamond edges broaden, until about 150~K where thermal broadening is sufficient to blur the diamond shapes. The large addition energies in our system limit the number of diamonds visible in the experimentally accessible gate range. For some devices, only a lower boundary on the addition energy can be set (for instance, about 300~meV for the right diamond in device E).

\section{Theoretical modeling}

\begin{figure*}[ht!]
\centering \includegraphics[width=150mm]{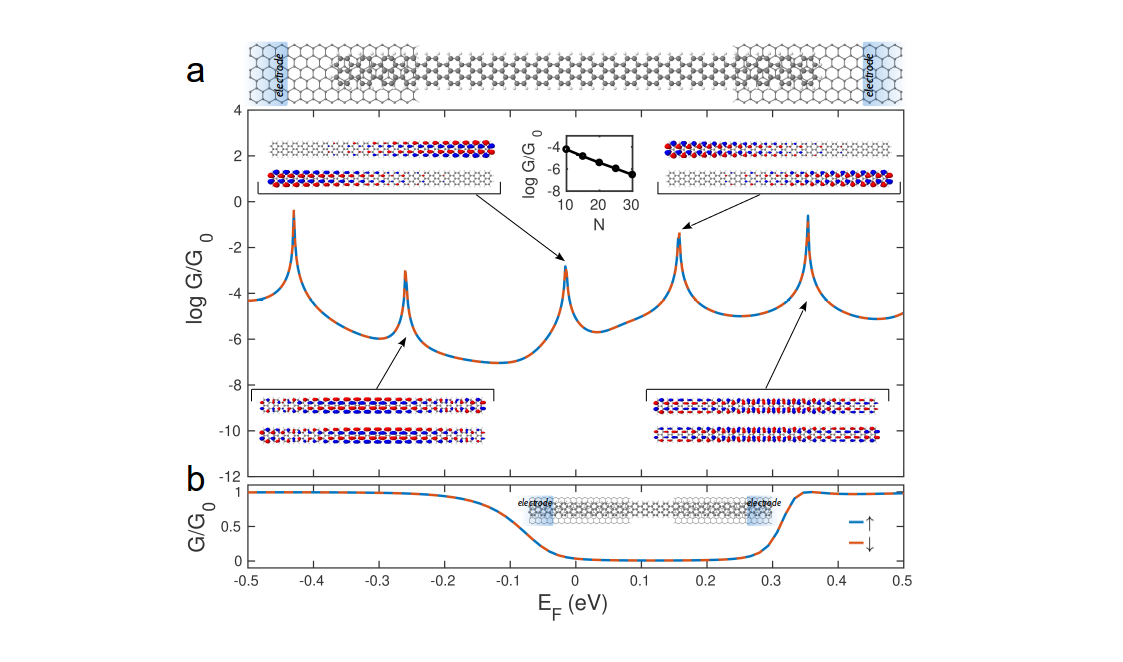}
\caption{ \textbf{Electronic properties of 5-AGNR/graphene nanodevices.} a) Structure and computed conductance of the nanodevice consisting of a 5-AGNR bridging a graphene nanogap. The blue areas at both sides of the junction represent the electrodes. The insets give the results of the computed wave functions for the majority and minority spins and the conductance versus the number of fused naphtalene units (N) while keeping the GNR/electrode overlap constant. b) Electrical conductance of an infinite 5-AGNR bridging a graphene nanogap. }
\label{theory}
\end{figure*}

To gain a deeper understanding of the charge-transport properties of our devices, we simulate the behavior of a 5-AGNR between two graphene electrodes, as depicted in Fig.~\ref{theory}a. For such a device, we obtain the mean-field Hamiltonian from density functional theory (DFT) and use the Gollum\cite{gollum,nanotech18} quantum transport code to compute the electron transmission through the 5-AGNR from the left graphene electrode to the right one (more details are provided in the methods section). Note that the size of the graphene nanogap is a few nanometers in our experiments; we therefore expect a phase coherent transport regime. 

Fig.~\ref{theory}a presents the calculated zero-bias conductance curve, alongside the wave functions corresponding to the resonances closest to the Fermi energy. The plot shows that these two resonances correspond to orbitals localized at the two extremities of the 5-AGNR. These end states are located closer to the Fermi energy than the resonances resulting from the fully delocalized 5-AGNR bulk states, located around -0.25 eV and +0.35 eV. Those resonances can be identified with the highest occupied molecular orbital level (HOMO) and lowest unoccupied molecular orbital level (LUMO) of the infinite ribbon. Such states have recently been observed for GNRs using scanning tunneling microscopy \cite{Schull}. As comparison, Fig.~\ref{theory}b presents the transmission of an infinite 5-AGNR bridging the two graphene electrodes. Here, the 5-AGNR is considered to be infinitely long and no end states are present. The figure shows that the resulting bandgap is about 0.4 eV, which is smaller than the HOMO-LUMO gap of the finite-length 5-AGNR. The reduction of the finite-ribbon gap is a result of the emergence of the end states (-20 meV and + 150 meV), yielding transport channels within the bandgap of the infinite ribbon. As a result, although counterintuitive, the transport gap of a finite-sized 5-AGNR is smaller than that of an infinitely long ribbon. This decrease in the HOMO-LUMO gap is accompanied by an increase of the electrical conductance. 

Moreover, as the size of the nanogap cannot be accurately determined, we performed DFT calculations to account for the variability of the 5-AGNR length between the graphene electrodes. We calculated the length-dependent conductance by increasing the number of fused naphthalene units within the nanogap of the device in Fig.~\ref{theory}b and S4b. We obtained a low decay factor of ca. 0.6 nm$^{-1}$ where the conductance only decreases by 2 orders of magnitude upon an increase of the length of the 5-AGNR from 3.5 to 14 nm (inset of Fig.~\ref{theory}a). The transmission function for such devices is shown in Fig.~S4 of the Supporting Information.

\section{Discussion}

The room-temperature measurements of the 5-AGNR devices show a semi-metallic behavior with linear IVs, albeit with high resistances. These high resistances may be due to a weak coupling between the GNRs and the graphene electrodes. This is in agreement with our low-temperature measurements, which exhibit the signatures of weakly-coupled quantum dots. This weak coupling is attributed to the presence of tunneling barriers at the graphene/GNR interfaces, which may be due to the presence of defects in the graphene and/or polymer residues from the transfer process, resulting in a reduced orbital overlap.

The observed sample-to-sample variations may also be related to differences in length of the ribbons and the number of ribbons bridging the gap. In this study, the average length of the GNRs varies between 2 and 10 nm. DFT calculations predict that the HOMO-LUMO gap of a 5-AGNR depends on the ribbon length, in particular for short ribbons, a trend which has been confirmed experimentally using STM measurements\cite{ch7_metalic5}. We note that the addition energies extracted from the different junctions are comparable to the 5-AGNR HOMO-LUMO gap value obtained from DFT calculations, which we, however, would expect to underestimate the bandgap\cite{Okan2017}. A quantitative comparison between experiment and theory therefore remains elusive, requiring GW calculations including advanced image-charge correction models in order to access the energy levels of the substrate-supported GNRs\cite{Okan2017,Talirz2017}. Nevertheless, our calculations qualitatively reproduce the decrease of the bandgap with increasing ribbon length.

A noteworthy result from the DFT calculations is that the localized states formed at the end of the short ribbons are predicted to be spin-polarized. A more elaborate description of this effect is presented in Fig. S3 and S4 of the supporting information. To confirm these spin-polarized states, measurements using spin-polarized currents under magnetic field are required, which are beyond the scope of this paper. This prediction nonethless offers interesting prospects for 5-AGNR-based spintronic devices.

Finally, we note that the interaction of the 5-AGNRs with the graphene electrodes may result in additional features in the SET regime. In particular, quantum interference effects within the graphene leads have been shown to give rise to an energy-dependent transmission and fluctuations in the sequential tunnel-rates\cite{ch7_pascal}. However, such excitations originating from the leads are not parallel to the end of the coulomb diamonds, in contrast to excitations coming from the channel itself. This leads us to conclude that this scenario can be excluded in our measurements.

\section{Conclusion}

To conclude, we have demonstrated the successful integration of short 5-AGNRs in a device geometry using sub-5 nm graphene junctions. At room temperature, we observe linear, metal-like IV-curves. This indicates that the bandgap of the 5-AGNR is very small, which qualitatively agrees with DFT calculations and STM measurements. At 13~K, we observe single-electron transistor behavior, with addition energies reaching a few hundred meV. Our first-principles transport calculations attribute this small energy gap to the finite size of the ribbons, and the resulting presence of end states at the two termini of the ribbon. Finally, the observation of additional resonances in the SET regime with energies around 70 meV points supports the fact that transport through the junction in a particular gate voltage range is dominated by a single quantum dot. This validates the use of graphene electrodes to contact ultra-small GNRs as well as the prospects of GNR-based electronic devices.

\section{Methods}

\begin{small}

\textbf{GNR synthesis:} In a first step, the precursor molecule which consists of an isomeric mixture of 3,9-dibromoperylene and 3,10-dibromoperylene (DBP) was sublimated at $~160ºC$ onto a Au(788) surface kept at room temperature \cite{Raman-GNR}. A slow annealing process ($~0.2ºC/s$) up to $~225ºC$ allowed the formation of 5-AGNRs via polymerization and cyclodehydrogenation. 

\textbf{Device fabrication:} The starting point of the devices is a doped (p++) silicon chip with 285 nm of thermally grown SiO$_2$. The CVD-grown graphene is transferred onto the silicon chip using a PMMA wet transfer technique. The nanogaps are then made in two steps: first e-beam lithography is used to pattern the graphene films into 400 nm wide stripes (combined with reactive ion etching) and metallic contacts (5 nm~Ti/40 nm Au deposited using e-beam evaporation). In a second step, nanogaps are formed in the graphene stripes using the EB technique\cite{}.

\textbf{Transfer of the GNRs:} This transfer method preserves both the structural quality and uniaxial alignment of the ribbons, as demonstrated by the Raman results in Fig.~\ref{figure2}. 
In a first step toward the electrochemical delamination of the 5-AGNRs from the Au (788) surface, poly(methylmethacrylate) (PMMA) was spin-coated on the 5-AGNR covered Au surface to be used as a support layer during the ribbon transfer. An aqueous solution of NaOH (1M) was employed as electrolyte in the electrochemical process and a DC-voltage was applied between the PMMA/5-AGNR/Au (788) cathode and a glassy carbon electrode used as anode. During this process, water undergoes reduction, resulting in hydrogen bubbles emerging at the 5-AGNR/Au interface. The $H_2$ bubbles provide enough force to detach the 5-AGNR film from the Au surface, starting from the edges and followed by the permeation of the electrolyte solution into the interface as the edges delaminate. The 5-AGNR/PMMA film is then transferred onto the graphene-nanogap devices using a wet-transfer technique. After transfer, the sample undergoes a two-step annealing in order to increase the contact between the GNRs and the graphene-nanogap device. First, the sample is heated to $\approx$80$^\circ$C for 10 minutes followed by a second annealing at $\approx$110$^\circ$C for 20 minutes. Finally PMMA is dissolved in hot acetone for 15 minutes.

\textbf{Raman spectroscopy:} Raman spectra were acquired with a WITec confocal Raman microscope (WITec Alpha 300R) equipped with a home-built vacuum chamber. Spectra were acquired with 20 mW laser power on the metal contact pad next to the junction B to exclude background from the silicon substrate. Time-series measurements show no radiation damage over the acquisition time. Figure 2a is a spatial average from 266 individual points after subtraction of a polynomial background.

\textbf{Density functional theory calculations:} The optimized geometry and ground state Hamiltonian and overlap matrix elements of each structure studied in this paper were self-consistently obtained using the SIESTA \cite{soler2002siesta} implementation of the density functional theory (DFT). SIESTA employs norm-conserving pseudo-potentials to account for the core electrons and linear combinations of atomic orbitals (LCAO) to construct the valence states. The generalized gradient approximation (GGA) of the exchange and correlation functional is used with the Perdew-Burke-Ernzerhof (PBE) parameterization and a double-$\zeta$ polarized (DZP) basis set. The real-space grid is defined with an equivalent energy cut-off of 250 Ry. The geometry optimization for each structure is performed to the forces smaller than 20 meV/$\AA$. For the band structure calculation, the unit cell of 5-AGNR (inset of Fig.~\ref{theory}a) was sampled by a $1 \times 1 \times 500$ Monkhorst-Pack k-point grid.

\textbf{Transport calculations:} The mean-field Hamiltonian obtained from the converged SIESTA DFT calculation was combined with the quantum transport code Gollum \cite{gollum,nanotech18}, to calculate the phase-coherent, elastic scattering properties of the each system consisting of left (source) and right (drain) graphene leads connected to the scattering region formed from 5-A GNR bridge (top panel of Fig.~\ref{theory}b). The transmission coefficient $T(E)$ for electrons of energy E (passing from the source to the drain) is calculated via the relation $T(E)=trace(\Gamma_R(E)G^R(E)\Gamma_LG^{R\dagger}(E))$. In this expression, $\Gamma_{L,R} = i(\Sigma_{L,R}(E)-\Sigma_{L,R}^\dagger(E))$ describe the level broadening due to the coupling between left (L) and right (R) electrodes and the central scattering region, $\Sigma_{L,R}(E)$ are the retarded self-energies associated with this coupling and $G^R=(ES-H-\Sigma_L-\Sigma_R)^{-1}$ is the retarded Green’s function, where H is the Hamiltonian and S is the overlap matrix. From the resulting transmission coefficient, the conductance is calculated by Landauer formula $G=G_0\int dE T(E)(-\partial f(E,T)\partial E)$ where $G_0=2e^2/h$ is the conductance quantum, $f(E,T)=(1+exp((E-E_F)/k_BT)^{-1}$ is the Fermi-Dirac distribution function, T is the temperature and $k_B= 8.6\time10^{-5}$eV/K is Boltzmann’s constant.

\textbf{Edge state:} 
DFT calculations for finite-length ribbons (different numbers of fused naphthalene units) show that while short ribbons up to 7 naphthalene units are non-magnetic, for longer ribbons the ground state spin configuration is antiferomagnetic (AF) (see Fig. S3a and spin density calculations of Fig. S3b,c of the SI). We also found that in ribbons with more than 18 naphtalene units, the energy differences between antiferromagnetic and ferromagnetic spin arrangements is less than room temperature (see Fig. S4d of the SI). 
\end{small}

\section{Data and code availability}
All data and measurements/analysis codes used in this study are available upon reasonable request.
\section{Acknowledgments}

This work was supported by the EC FP7-ITN MOLESCO grant (no. 606728) and the FET open project QuIET (no. 767187). This work was supported by EPSRC grants EP/P027156/1, EP/N03337X/1 and EP/N017188/1. M.P. acknowledges funding by the EMPAPOSTDOCS-II program which is financed by the European Unions Horizon 2020 research and innovation program under the Marie Sk\l{}odowska-Curie grant agreement number 754364. H.S. acknowledges the UKRI for Future Leaders Fellowship no. MR/S015329/1 and Leverhulme Trust for Early Career Fellowship no. ECF-2017-186. S.S. acknowledges the Leverhulme Trust for Leverhulme Early Career Fellowship No. ECF-2018-375. O.B. acknowledges technical support from the Binning and Rohrer Nanotechnology Center (BRNC), Ruschlikon, in particular from Antonis Olziersky for the e-beam lithography. G.B.B, P.R. and R.F. acknowledge funding by the Swiss National Science Foundation under Grant No 20PC21-155644, the European Unions Horizon 2020 research and innovation program under grant agreement number 785219 (Graphene Flagship Core 2), and the Office of Naval Research BRC Program under the grant N00014-12-1-1009.

\section{Author contribution}
O.B. and M.E. fabricated the samples. 
M.E performed the electrical characterisation. 
J.O. performed the Raman characterisation.
T.P., A.N. and K.M synthesized the precursor molecule.
G.B.B and Q.S. synthesized the 5-AGNR.
M.E and M.P. performed the data analysis. 
S.S and H.S. performed the DFT calculations.
P.R., H.S, C.L., R.F., and M.C. supervised the study.
M.E, M.P., H.S., G.B.B. and M.C. wrote the manuscript.
All authors participated in the discussion of the results and commented on the manuscript.

\bibliography{biblio}
\bibliographystyle{naturemag}

\end{document}